\documentstyle[psfig,aps,prl,amssymb]{revtex}

\begin{document}
\title{Bures Geometry of the
Three-Level Quantum Systems. I}
\author{Paul B. Slater}
\address{ISBER, University of California, Santa Barbara, CA 93106-2150\\
e-mail: slater@itp.ucsb.edu, FAX: (805) 893-7995}

\date{\today}

\draft

\maketitle

\vskip -0.1cm

\begin{abstract}
We compute --- using a formula of Dittmann --- the Bures metric tensor
($g$) for the eight-dimensional state space  of three-level quantum systems, 
employing a newly-developed Euler angle-based 
parameterization of the $3 \times 3$ 
 density matrices. Most of the individual metric elements ($g_{ij}$) are
found to be expressible in relatively compact form, many of them in fact
being exactly {\it zero}.
\end{abstract}

\vspace{.2cm}
\hspace{1.5cm} Keywords: Bures metric, three-level quantum systems, spin-1 
systems, density matrices, 

\hspace{1.5cm} orthogonal parameters, Euler angles, unitary transformations

\vspace{.15cm}

\hspace{1.5cm} Mathematics Subject Classification (2000): 81Q70, 53Axx

\tableofcontents

\vspace{.1cm}
\section{INTRODUCTION}

The Bures metric is a distinguished member --- the {\it minimal}
 one --- of the
(nondenumerable) 
family of {\it monotone} metrics on the quantum systems \cite{petzsudar}.
Its contemporary 
study was pioneered by Armin Uhlmann \cite{uhl1,uhl2}, along with several
of his associates at the University of Leipzig 
\cite{hub1,ditt1,ditt2,ditt3,ditt4}. 
In particular, Jochen Dittmann has derived several {\it explicit} formulas
(ones not requiring knowledge of the eigenvalues of density matrices) for the
Bures metric \cite{ditt3,ditt4}. Slater \cite{slat1} --- interpreting 
the {\it volume element}
of the metric as a natural (unnormalized) measure on the 
quantum systems --- applied this work to
certain low-dimensional subsets of the fifteen-dimensional set of
$4 \times 4$ density matrices to obtain ``exact Bures probabilities that
two quantum bits are classically correlated'' (cf. \cite{SLAT,ZHSL}).

The Bures metric on the three-dimensional convex set of the $2 \times 2$ 
density matrices (making use of Cartesian coordinates ($x,y,z$)),
\begin{equation} \label{2by2}
\rho ={1 \over 2} 
\pmatrix{1 + z & x + i y \cr x - i y & 1-z \cr}, \qquad (0 \leq x^2+y^2+z^2 
\leq 1)
\end{equation}
has been intensively studied. 
The corresponding metric tensor
\begin{equation} \label{hhe}
g = {1 \over 4 (1-x^2-y^2-z^2)} \pmatrix{1-y^2-z^2 & x y & x z \cr
x y & 1-x^2 -z^2 & yz \cr
x z & y z & 1-x^2-y^2 \cr}
\end{equation}
can be obtained by application of an (early) formula of Dittmann
\cite[eq. (3.7)]{ditt3},
\begin{equation} \label{fu}
d_{Bures}(\rho,\rho+ \mbox{d} \rho)^{2} = {1 \over 4} \mbox{Tr} \lbrace
\mbox{d} \rho \mbox{d} \rho + {1 \over |\rho|} (\mbox{d} \rho -\rho
\mbox{d} \rho) (\mbox{d} \rho - \rho \mbox{d} \rho) \rbrace.
\end{equation}
In spherical coordinates ($r,\theta,\phi$) the tensor (\ref{hhe})
takes a {\it diagonal} form
\begin{equation} \label{DIAG}
g = {1 \over 4} \pmatrix{{1 \over (1-r^2)} & 0 & 0 \cr 0 & r^2 & 0 \cr
0 & 0 & r^2 \sin^{2}{\theta} \cr}.
\end{equation}

The Bures metric can be viewed as the
standard metric on the surface of a three-sphere 
\cite{hub1,bm}. As such, Hall \cite[p. 128]{hall} has written 
that ``the Bures metric for
a two-dimensional system corresponds to the surface of a unit four-ball,
\linebreak
i. e., to the maximally symmetric three-dimensional space of positive
curvature (and may be recognized as the spatial part of the Robertson-Walker
metric in general relativity). This space is homogeneous and isotropic, and hence the Bures metric does not distinguish a preferred location or direction in
the space of density operators. Indeed, as well as rotational symmetry in
Bloch coordinates (corresponding to unitary invariance), the metric has a
further set of symmetries generated by the infinitesimal transformations
\begin{equation}
r \rightarrow r + \epsilon (1-r^2)^{1 \over 2} a,
\end{equation}
(where $a$ is an arbitrary three-vector [and $r$, radial distance in the
Bloch sphere of two-level quantum systems \cite{bm}]).'' 
Petz and Sud\'ar  observed that in ``the case of the [Bures] metric 
of the symmetric logarithmic derivative the tangential component is
independent of $r$'' \cite[p. 2667]{petzsudar}.

A principal goal of the present study is to determine any such symmetries
possessed by the Bures metric when one proceeds from the study of the
two-level quantum systems to that of the three-level quantum systems. 
One should be aware, though, that
Dittmann has noted that in this case, the space ``is not a space of constant
curvature and not even a locally symmetric space, in contrast to 
what the case 
of two-dimensional density matrices might suggest'' \cite{ditt3}. 
(In a locally symmetric space, the sectional curvature is invariant under 
parallel displacement, and the covariant derivative of the curvature
tensor field vanishes \cite{ditt1,helgason}. A formula for the scalar 
curvature of the monotone metrics for general $n$-level quantum systems 
is given in \cite{ditt11}, cf. \cite{ditt12}.) In other
work \cite{dittym}, Dittmann has shown that the gauge field 
defining the Bures metric satisfies
the source-free Yang-Mills equation. Petz 
\cite[Thm. 3.4]{petzmore} has established
that the Bures metric is the only monotone metric that is both
``Fisher adjusted'' and ``Fubini-Study adjusted''.
\section{METHODOLOGY}

Slater \cite{slat2} (cf. \cite[eqs. (6), (7)]{slat1}) applied a
 formula (cf. (\ref{fu})) 
of Dittmann \cite[eq. (3.8)]{ditt3} for the specific case
of the three-level quantum systems,
\begin{equation} \label{form1}
g^{B}_{\rho} = {1 \over 4} \mbox{Tr} \lbrace  \mbox{d} \rho \mbox{d} \rho  +
{3 \over 1 - \mbox{Tr} \rho^3}
 (\mbox{d} \rho -\rho \mbox{d} \rho ) (\mbox{d} \rho -\rho 
\mbox{d} \rho) +{3 |\rho| \over 1 - \mbox{Tr} \rho^{3} } 
(\mbox{d} \rho  - \rho^{-1} \mbox{d} \rho) (\mbox{d} \rho -\rho^{-1} 
\mbox{d} \rho ) \rbrace,
\end{equation}
to the particular instance (a simple extension of the 
two-level quantum systems (\ref{2by2})) of a {\it four}-dimensional subset,
\begin{equation}
\rho = {1 \over 2} \pmatrix{v+z & 0 & x- i y  
\cr 0 & 2 -2 v & 0 \cr x + i y & 0 & v -z \cr},
\end{equation}
of the eight-dimensional convex set of $3 \times 3$ density matrices
\cite{bloore}. Now, 
in the present study, we apply this same
 formula (\ref{form1}) to the {\it full}
eight-dimensional convex set of the three-level quantum 
systems itself. Of crucial 
and central importance
here will be the use of a 
recently-developed ``Euler angle'' parameterization of
these density matrices \cite{bb,us}. In this parameterization, one takes an
arbitrary density matrix ($\rho$) to be 
expressed in the (``Schur/Schatten'') form \cite[sec. 3]{ditt4}  
\cite[p. 3725]{twam} 
\cite[p. 53]{hasegawa}
\begin{equation} \label{pte}
\rho = U \rho^{'} U^{\dagger}.
\end{equation}
Here
\begin{equation}
U= \mbox{e}^{i \lambda_{3} \alpha} \mbox{e}^{i \lambda_{2} \beta}
\mbox{e}^{i \lambda_{3} \gamma} \mbox{e}^{i \lambda_{5} \theta}
\mbox{e}^{i \lambda_{3} a} \mbox{e}^{i \lambda_{2} b},
\end{equation}
is a member of $SU(3)$, the three immediately relevant 
(of the eight) Gell-Mann matrices \cite{po}
being
\begin{equation}
\lambda_{2} = \pmatrix{0 & - i & 0 \cr i & 0 & 0 \cr 0 & 0 & 0 \cr},\qquad
\lambda_{3} = \pmatrix{1 & 0 & 0 \cr
0 & -1 & 0 \cr 0 & 0 & 0 \cr}, \qquad \lambda_{5} = \pmatrix{0 & 0 & -i \cr
0 & 0 & 0 \cr i & 0 & 0 \cr}.
\end{equation}
Making use of spherical coordinates ($\theta_{1}, \theta_{2}$),
\begin{equation}
\rho^{'} = \pmatrix{\cos^2{\theta_{1}} & 0 & 0 \cr
0 & \sin^{2}{\theta_{1}} \cos^{2}{\theta_{2}} & 0 \cr
0 & 0 & \sin^{2}{\theta_{1}} \sin^{2}{\theta_{2}} \cr}.
\end{equation}
An appropriate set of ranges of the eight angles (by which all the 
$3 \times 3$ density matrices can be reproduced without duplication) is 
\cite[eqs. (11), (12)]{bb}
\begin{equation}
0 \leq \alpha, 
\gamma, a \leq \pi, \quad 0 \leq \beta, 
\theta, b \leq {\pi \over 2}, \quad 0 \leq \theta_{1} \leq 
\cos^{-1}{1 \over \sqrt{3}}, \quad 0 \leq \theta_{2} \leq {\pi \over 4}.
\end{equation}

We have inserted the so-parameterized 
$3 \times 3$ density matrix (\ref{pte}) into 
formula (\ref{form1}) to obtain the $8 \times 8$ Bures metric tensor.
Since, by construction, we have explicit knowledge of the eigenvalues
($\lambda$'s)
and eigenvectors of $\rho$, we could alternatively have directly
applied the {\it general} formula
for the Bures metric in the $n$-dimensional case \cite[eq. (10)]{hub1},
\begin{equation}
d_{Bures}(\rho,\rho + \mbox{d} \rho)^{2} = 
{1 \over 2} \sum_{i,j =1}^{n} {|<i|\mbox{d} \rho|j>|^{2} \over \lambda_{i} 
+ \lambda_{j}},
\end{equation} 
or that given by Proposition 4 in \cite{ditt4}.

\section{ELEMENTS OF THE BURES METRIC TENSOR}

Initially, all the entries of the tensor computed using 
(\ref{form1}) --- implemented in MATHEMATICA --- were given 
by extremely large
complicated expressions. However, in a number of cases, both through exact
computations and numerical experimentation, we were able to arrive
at certain relatively compact (if not simply strictly 
{\it zero} themselves) expressions for the individual metric elements.

The first remarkable item to note is that (as repeated numerical 
experiments indicate) {\it all} the entries of the tensor
are {\it independent} of the Euler angle $\alpha$. 
Further numerical investigations have convinced us that
many of the entries of the tensor are, in fact, zero (cf. \cite{tod,cox}). 
(In the case of the two-level quantum systems, the off-diagonal 
entries of the Bures metric tensor 
(\ref{DIAG}) are zero, if spherical --- as 
opposed to Cartesian --- coordinates
are employed.)
For example,
the  
spherical coordinates $\theta_{1}$ and $\theta_{2}$ are both
orthogonal to the other seven coordinates. The diagonal entry 
($g_{\theta_{1} \theta_{1}}$) of the Bures
metric tensor ($g$) 
corresponding to the pairing $(\theta_{1}, \theta_{1})$ is
simply 1, while the diagonal entry 
($g_{\theta_{2} \theta_{2}}$) corresponding to the pairing
$(\theta_{2},\theta_{2})$ is $\sin^{2}{\theta_{1}}$.

Let us summarize our present state of 
explicit knowledge regarding the
Bures metric elements ($g_{ij}$) for the three-level quantum  systems.
We write the corresponding (symmetric) matrix, using the ordering
of coordinates (and hence rows and columns) 
\begin{equation}
(\alpha, \gamma, a, \beta,b,\theta,\theta_{1},\theta_{2})
\end{equation}
 as
\begin{equation} \label{gmatrix}
g =\pmatrix{? & ? & g_{13} & ? & g_{15} & g_{16} & 0 & 0 \cr
\cdot & g_{22} & g_{23} & g_{24} & 0 & 0 & 0 & 0 \cr
\cdot & \cdot & g_{33} & g_{34} & 0 & 0 & 0 & 0 \cr
\cdot & \cdot & \cdot & g_{44} & g_{45} & g_{46} & 0 & 0 \cr
\cdot & \cdot & \cdot & \cdot & g_{55} & 0 & 0 & 0 \cr
\cdot & \cdot & \cdot & \cdot & \cdot & g_{66} & 0 & 0 \cr
\cdot & \cdot & \cdot & \cdot & \cdot & \cdot & 1 & 0 \cr
\cdot & \cdot & \cdot & \cdot & \cdot & \cdot & \cdot & 
\sin^{2}{\theta_{1}} \cr}.
\end{equation}
Our specific element-by-element results are now presented.
\subsection{$g_{55} = g_{b b}$}
\begin{figure} 
\centerline{\psfig{figure=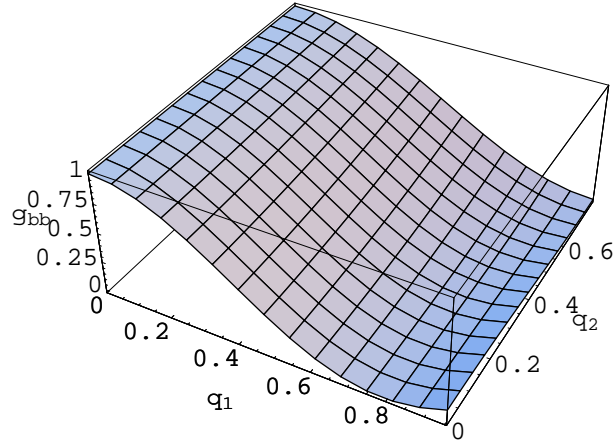}}
\caption{Diagonal (5,5)-entry, corresponding to the Euler angle 
$b$,  of the Bures metric tensor (\ref{gmatrix}) for the three-level quantum
 systems. This term --- which can be inverted (\ref{INVERT}) --- enters as 
well 
into many of the expressions for the other metric elements.}
\label{f1}
\end{figure}

We have (Fig.~\ref{f1})
\begin{equation} \label{g55eq}
g_{55}=g_{b b} ={t^{2}
 \over  16 u_{+}},
\end{equation}
where  (cf. \cite[eq. (28)]{slat3})
\begin{equation}
t = 2 + 6 \cos{2 \theta_{1}} + \cos{2(\theta_{1}-\theta_{2})} 
-2 \cos{2 \theta_{2}} +\cos{2 (\theta_{1}+\theta_{2})}
\end{equation}
and
\begin{equation}
u_{\pm}=   3 + \cos{2 \theta_{1}} \pm 
2 \cos{2 \theta_{2}} \sin^{2}{\theta_{1}}.
\end{equation}
\subsection{$g_{13}=g_{\alpha a}$}
\begin{equation}
g_{13} =g_{\alpha a} = {g_{55} \over 4} \lbrace (3 + \cos{2 \theta}) \cos{2 \beta}
\sin^{2}{2 b} + 2 \cos{2 (a + \gamma)} \cos{\theta} \sin{4 b} \sin{2 \beta}
\rbrace.
\end{equation}
\subsection{$g_{15} = g_{\alpha b}$}
\begin{equation}
g_{15} = g_{\alpha b} = g_{55} \cos{\theta} \sin{2 \beta} 
\sin{2 (a + \gamma)}.
\end{equation}
\subsection{$g_{16}=g_{\alpha \theta}$}
\begin{figure}
\centerline{\psfig{figure=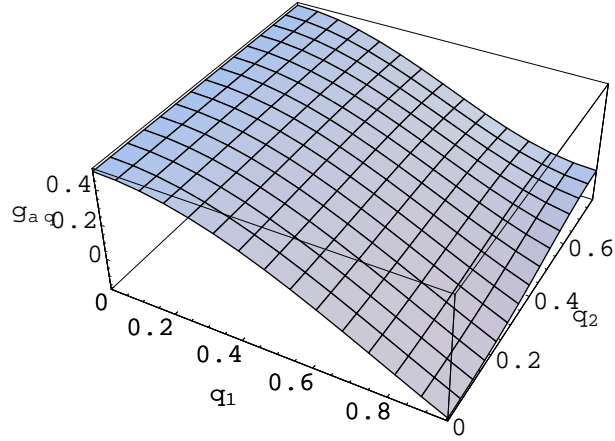}}
\caption{Euler angle-independent factor of metric element corresponding to
the (1,6)-entry of (\ref{gmatrix})}
\label{kdy}
\end{figure}

\begin{equation}
g_{16} = g_{\alpha \theta} =  {v \over 32 u_{-}} \sin{2 b} \sin{2 \beta} 
\sin{2 (a + \gamma)} \sin{\theta},
\end{equation}
where
\begin{equation}
v=  15 + 28 \cos{2 \theta_{1}} + 21 \cos{4 \theta_{1}} + 
4 (7 + 9 \cos{2 \theta_{1}}) \cos{2 \theta_{2}} \sin^{2}{\theta_{1}} \quad -
\end{equation}
\begin{displaymath}
 - \quad 
4 (5 + 3 \cos{2 \theta_{1}}) \cos{4 \theta_{2}} \sin^{2}{\theta_{1}} + 8 
\cos{6 \theta_{2}} \sin^{4}{\theta_{1}} .
\end{displaymath}
In Fig.~\ref{kdy} we plot the Euler angle-independent part of $g_{16}$, 
that is ${v \over 32 u_{-}}$.
\subsection{$g_{22} = g_{\gamma \gamma}$}
\begin{equation}
g_{22} =g_{\gamma \gamma} = {1 \over 16 \kappa} \lbrace -g_{55} \kappa 
\cos^{4}{b} (3 + \cos{2 \theta})^{2} + 4 \cos^{2}{b} (g_{55} \kappa +
\mu \cos^{2}{\theta} \quad  +
\end{equation}
\begin{displaymath}
+ \quad  4 (\kappa + \upsilon) \cos^{2}{2 \theta_{2}} 
\cos^{4}{\theta} \sin^{2}{\theta_{1}}) + 16 \kappa \cos^{2}{2 \theta_{2}}
 \cos^{2}{\theta} \sin^{2}{\theta_{1}} \sin^{2}{\theta} \rbrace,
\end{displaymath}
where 
\begin{equation}
\kappa = 35 + 28 \cos{2 \theta_{1}} + \cos{4 \theta_{1}} - 
8 \cos{4 \theta_{2}} \sin^{4}{\theta_{1}}
\end{equation}
\begin{equation}
\upsilon = -4 (1 +3 \cos{2 \theta_{1}}) (7 +5 \cos{2 \theta_{1}}) 
\sec{2 \theta_{2}} -16 \cos{2 \theta_{2}} \sin^{4}{\theta_{1}},
\end{equation}
and
\begin{equation}
\mu= -\sin^{2}{\theta_{1}} \lbrace 1621 + 125 \cos{2 \theta_{2}} + 46 \cos{4 \theta_{2}} + 4 \cos{2 \theta_{1}} (261 + 49 \cos{2 \theta_{2}} + 10 \cos{4 \theta_{2}}) \quad + 
\end{equation}
\begin{displaymath}
+ \quad \cos{4 \theta_{1}} (151 + 63 \cos{2 \theta_{2}} + 42 \cos{4 \theta_{2}})
-768 \csc^{2}{\theta_{1}} + 8 (\cos{6 \theta_{2}} -\cos{8 \theta_{2}}) 
\sin^{4}{\theta_{1}} \rbrace.
\end{displaymath}
\subsection{$g_{23} = g_{\gamma a}$}
\begin{equation} \label{g23}
g_{23} =g_{\gamma a} = {g_{55} \over  4 } (3 +\cos{2 \theta}) \sin^{2}{2 b}.
\end{equation}
\subsection{$g_{24}= g_{\gamma \beta}$}
\begin{equation}
g_{24}= g_{\gamma \beta} = {3 t \cos{\theta} \sin{2 b} \sin{2 (a + \gamma)} 
\sin^{2}{\theta_{1}} (\cos{2 \theta} - i \sin{2 \theta}) \over
256 (-1 +\cos^{6}{\theta_{1}} +\sin^{6}{\theta_{1}} (\cos^{6}{\theta_{2}} 
+\sin^{6}{\theta_{2}}))}
\end{equation}
\begin{displaymath}
\cos^{2}{\theta_{1}} (-p \cos{2 b} (-1 + 4 \cos{2 \theta_{2}} + 
\cos{4 \theta_{2}}) +q (-1 + 7 \cos{2 \theta_{2}} - 3 \cos{4 \theta_{2}}
+\cos{6 \theta_{2}})) \qquad +
\end{displaymath}
\begin{displaymath}
+\quad 2 \cos^{4}{\theta_{1}} (p  \cos{2 b} \cos^{2}{\theta_{2}} 
(3 + \cos{2 \theta_{2}}) + q (4 -3 \cos{2 \theta_{2}} + \cos{4 \theta_{2}})
\sin^{2}{\theta_{2}}) + ( - p \cos{2 b} + q (-1 +2 \cos{2 \theta_{2}})) 
\sin^{2}{2 \theta_{2}},
\end{displaymath}
where
\begin{displaymath}
p= 1 + 6 e^{2 i \theta} +e^{4 i \theta}, \quad q= (-1+e^{2 i \theta})^{2}.
\end{displaymath}
\subsection{$g_{33}=g_{a a}$}
\begin{equation}
g_{33} = g_{a a} = g_{55} \sin^{2} {2 b} .
\end{equation}
\subsection{$g_{34} = g_{a \beta}$}
\begin{equation}
g_{34} =g_{a \beta}= -{g_{55} \over 2 } \cos{\theta} 
\sin{4 b} \sin{2 (a + \gamma)}.
\end{equation}
\subsection{$g_{44} = g_{\beta \beta}$}
\begin{equation} 
g_{44} =g_{\beta \beta} = - g_{55} \cos^{2}{b} \cos^{2}{\theta} \sin^{2}{b}
\sin^{2}{2(a + \gamma)} + {\zeta \over 32 \kappa}, 
\end{equation}
where
\begin{equation}
\zeta = -\csc^{2}{\theta_{1}} \lbrace -101 + 
12 \cos{4 \theta_{1}} + 64 \cos{6 \theta_{1}}
+25 \cos{8 \theta_{1}} + 16 (61 + 100 \cos{2 \theta_{1}} + 31 \cos{4 \theta_{1}})
\end{equation} 
\begin{displaymath}
\cos{2 \theta_{2}} \cos{2 \theta} \sin^{4}{\theta_{1}} - 64 (5 + 7 \cos{2 \theta_{1}}) \cos{4 \theta_{2}} \sin^{6}{\theta_{1}} \quad +
\end{displaymath}
\begin{displaymath}
+ \quad 128 \cos{2 \theta_{2}} \cos{4 \theta_{2}} \cos{2 \theta} 
\sin^{8}{\theta_{1}} + 2 \cos^{2}{b} \sin^{2}{\theta_{1}} 
( 242 + 445 \cos{2 \theta_{1}} + 286 
\cos{4 \theta_{1}} \quad +
\end{displaymath}
\begin{displaymath}
 \quad 51 \cos{6 \theta_{1}} + 4 ((125 + 196 \cos{2 \theta_{1}}
+ 63 \cos{4 \theta_{1}}) \cos{2 \theta_{2}} -2 (29 + 28 \cos{2 \theta_{1}} 
+ 7 \cos{4 \theta_{1}})
\end{displaymath}
\begin{displaymath}
 \cos{4 \theta_{2}}) \sin^{2}{\theta_{1}} + 
32 (\cos{6 \theta_{2}} + \cos{8 \theta_{2}})  \sin^{6}{\theta_{1}}) 
\sin^{2}{\theta} \rbrace.
\end{displaymath}
\subsection{$g_{45} = g_{\beta b}$}
\begin{equation} \label{g45}
g_{45} = g_{\beta b} = g_{55} \cos{2 (a + \gamma)} \cos{\theta}.
\end{equation}
\subsection{$g_{46} = g_{\beta \theta}$}
\begin{equation}
g_{46} =g_{\beta \theta} =t { \cos{2(a +\gamma)}
 \sin{2 b} (2 \cos{2 \theta_{1}} - (\cos{4 \theta_{2}} -
 3 \cos{2 \theta_{2}}) 
 \sin^{2}{\theta_{1}}) \sin{\theta} \over 8 u_{-}}.
\end{equation}
\subsection{$g_{66} = g_{\theta \theta}$}
\begin{equation}
g_{66} = g_{\theta \theta} = {32 \cos^{2}{b} \cos^{4}{\theta_{1}} \over 
6 + 2 \cos{2 \theta_{1}} +  +\cos{2 (\theta_{1}-\theta_{2})}
-2 \cos{2 \theta_{2}}  +\cos{2(\theta_{1}+\theta_{2})} } \quad + 
\end{equation}
\begin{displaymath}
+ \quad  {1 \over 4} \lbrace -2 - 
4 \cos{2 \theta_{1}} + (- \cos{2 \theta_{2}} +
\cos{4 \theta_{2}}) \sin^{2}{\theta_{1}} -\cos{2 b} (6 \cos^{2}{\theta_{1}}
+ (\cos{2 \theta_{2}} +\cos{4 \theta_{2}}) \sin^{2}{\theta_{1}}) \rbrace.
\end{displaymath}

Since the Euler angles $a$ and $\gamma$  seem only to appear in the
$g_{ij}$'s in the additive 
combination $a+ \gamma$, we conducted a reparameterization of the form
$\gamma= \tau -a$. Then, we found that the entries of the 
associated $8 \times 8$
Bures metric tensor 
(again computed using (\ref{form1})) 
were not only independent of $\alpha$, as before, but
also of the parameter $a$.
\section{CONCLUDING REMARKS}

We would like to express guarded optimism that, with sufficient expenditure
of computational resources 
and/or added 
ingenuity and insight, the question marks 
in (\ref{gmatrix}) can be effectively
removed, and one proceed with 
supplementary analyses, such as inversion of the Bures metric 
tensor, for purposes of statistical estimation 
\cite{gm} \cite[eq. (7)]{slat3} 
and computation of the volume element of the metric, that is the
``quantum Jeffreys' prior'' \cite{slat2,kwek}. 
Let us note here that the inverse of the Bures metric tensor (\ref{hhe})
for the {\it two}-level quantum systems
takes the particularly simple form
\begin{equation}
g^{-1} = 4 \pmatrix{1-x^2 & - x y & - x z \cr
-x y  & 1-y^2 & -y z \cr
-x z & - y z  & 1-z^2 \cr}.
\end{equation}

However, we have 
confirmed that the remaining not 
explicitly expressed  $g_{ij}$'s in (\ref{gmatrix}) are not simply
products of two independent functions, one of the Euler angles
($\alpha,\gamma,a,\beta,b,\theta$), and the other of the spherical angles
($\theta_{1},\theta_{2}$). These three yet (relatively compactly)
unexpressed elements (that is, $g_{11} = g_{\alpha \alpha}, 
g_{12} = g_{\alpha \gamma}$ and $g_{14} = g_{\alpha \beta}$) 
are independent only of 
$\alpha$, and not of the other seven parameters.
If we set $\beta=b=0$, then $g_{14} =0$ and both 
$g_{11}$ and $g_{12}$ reduce to (cf. \cite[eq. (28)]{slat3})
\begin{equation}
{(-2 -6 \cos{2 \theta_{1}} +\cos{2 (\theta_{1}-\theta_{2})} -2 
\cos{2 \theta_{2}} +\cos{2 (\theta_{1}+\theta_{2})})^{2} \sin^{2}{2 \theta} 
\over 64 (3 + \cos{2 \theta_{2}} -2 \cos{2 \theta_{2}} 
\sin^{2}{\theta_{1}})}.
\end{equation}
If we set $\beta=\theta=0$, on the other 
hand, then both $g_{11}$ and $g_{12}$ reduce to
$g_{33}$, that is $g_{55} \sin^{2}{2 b}$, while $g_{44}$ reduces to
$-g_{55} \sin{4 b} \sin{2 (a + \gamma)} /2$.

We also can not rule out the 
possibility that some of the
more complicated expressions we have presented here --- such as $g_{22}$ 
and $g_{44}$ --- have, in fact, considerably simpler forms than have 
so far been uncovered.
In addition to the transformation $\gamma = \tau -a$, which as we have
already noted renders {\it all} the elements of the Bures metric tensor
independent of $a$, as well as $\alpha$, another quite interesting
reparameterization would be 
based on the inversion of the relation (\ref{g55eq}), since the 
element $g_{55}$ itself enters directly into  the expressions 
for many of the other elements. That
is, one has
\begin{equation} \label{INVERT}
\theta_{2} =\sec^{-1}{{2 \sqrt{2} \sin{\theta_{1}} \over
\sqrt{ 4 + g_{55} + 4 \cos{2 \theta_{2}} + \sqrt{g_{55}} 
\sqrt{16 + g_{55} +16 \cos{2 \theta_{2}}}}}}.
\end{equation}
We have recomputed the Bures metric tensor, which we now denote 
$\tilde{g}$, again 
with (\ref{form1}), using $\tau$ and $g_{55}$ as
parameters, rather than $\gamma$ and $\theta_{2}$ as in our main 
analysis and, indeed, 
found  that
$\tilde{g}_{b b}$ has the expected form, that is  equalling 
$g_{55}$, and, similar type results for $\tilde{g}_{\alpha a}, 
\tilde{g}_{\alpha b}, 
\tilde{g}_{a a},\tilde{g}_{a b}$ and $\tilde{g}_{\beta b}$. Also, 
numerically  
$\tilde{g}_{\tau a}=g_{\gamma a}$. 

Since M. Byrd has indicated that he will shortly present an Euler angle
parameterization of $SU(4)$, parallel to that of
$SU(3)$ \cite{us} used here, it will, at that point, be of interest to 
similarly  attempt to
recreate the 
$15 \times 15$ 
Bures metric tensor for the {\it four}-level quantum systems --- which 
are capable of describing the state of a pair of {\it qubits} 
(cf. \cite{kus}). For this 
task, rather than (\ref{form1}), it will be necessary to use one of the 
other 
``explicit formulae for the Bures metric'' given by Dittmann in \cite{ditt4}.

In part II of this paper, which is in preparation, we intend to report
further progress  in the realization and simplification of 
formulas for the entries of the 
$8 \times 8$ Bures metric
tensor {\it and} of its inverse. These results will be 
{\it applied} to the
study of the curvature properties of the metric
(cf. \cite{bilge}), following upon the 
demonstration of Dittmann \cite{dittym} that the curvature tensor for the
Bures metric satisfies the Yang-Mills equation.
We will report additional highly interesting features of the curvature.

\acknowledgments

I would like to express appreciation to the Institute for
Theoretical Physics for computational support in this research.

\end{document}